\documentclass[12pt]{article}

\usepackage{amssymb,amsmath,amsfonts,amssymb}
\usepackage{graphics,graphicx,color,verbatim}

\def\@abssec#1{\vspace{.05in}\footnotesize \parindent .2in
{\bf #1. }\ignorespaces}
\graphicspath{{/EPSF/}{Figures/}}
\newtheorem{theorem}{Theorem}%[section]
\newtheorem{lemma}{Lemma}[section]

\def \Rm {\mathbb R}
\def \Nm {\mathbb N}

\newcommand{\eps}{\varepsilon}

\newcommand{\E}{\mathbb E}
\renewcommand{\P}{\mathbb P}
\newcommand{\dsum}{\displaystyle\sum}
\newcommand{\dint}{\displaystyle\int}
\newcommand{\dprod}{\displaystyle\prod}
\newcommand{\pdr}[2]{\dfrac{\partial{#1}}{\partial{#2}}}

\newcommand{\bx}{\mathbf x}

\newcommand{\ban}{{\bar n}}

\newcommand{\cout}[1]{}

\newcommand{\mk}{\mathfrak k}
\newcommand{\mM}{\mathfrak M}
\newcommand{\ms}{\mathfrak s}

\newcommand{\m}{{\mathfrak m}}
\newcommand{\n}{{\mathfrak n}}

\newcommand{\pp}{{\mathfrak p}}
\newcommand{\PP}{{\mathfrak P}}

\newcommand{\mB}{{\mathcal B}}
\newcommand{\mH}{{\mathcal H}} \newcommand{\mI}{{\mathcal I}}

 \renewcommand{\arraystretch}{1.5}

\title{Convergence to SPDEs in Stratonovich form}

\author{Guillaume Bal \thanks{Department of Applied Physics and 
        Applied Mathematics, Columbia University, 
        New York NY, 10027; gb2030@columbia.edu}}

\begin{document}
 
\maketitle

%\tableofcontents

\begin{abstract}
  We consider the perturbation of parabolic operators of the form
  $\partial_t+P(x,D)$ by large-amplitude highly oscillatory spatially
  dependent potentials modeled as Gaussian random fields. The
  amplitude of the potential is chosen so that the solution to the
  random equation is affected by the randomness at the leading order.
  We show that, when the dimension is smaller than the order of the
  elliptic pseudo-differential operator $P(x,D)$, the perturbed
  parabolic equation admits a solution given by a Duhamel expansion.
  Moreover, as the correlation length of the potential vanishes, we
  show that the latter solution converges in distribution to the
  solution of a stochastic parabolic equation with a multiplicative
  term that should be interpreted in the Stratonovich sense. The
  theory of mild solutions for such stochastic partial differential
  equations is developed.
  
  The behavior described above should be contrasted to the case of
  dimensions that are larger than or equal to the order of the
  elliptic pseudo-differential operator $P(x,D)$. In the latter case,
  the solution to the random equation converges strongly to the
  solution of a homogenized (deterministic) parabolic equation as is
  shown in the companion paper \cite{B-CMP-1-08}. The stochastic model
  is therefore valid only for sufficiently small space dimensions in
  this class of parabolic problems.
\end{abstract}
 
%\begin{AMS}
%\end{AMS}

\renewcommand{\thefootnote}{\fnsymbol{footnote}}
\renewcommand{\thefootnote}{\arabic{footnote}}

\renewcommand{\arraystretch}{1.1}

%\begin{keywords}
\paragraph{keywords:}
Partial differential equations with random coefficients, Stochastic
partial differential equations, Gaussian potential, iterated
Stratonovich integral, Wiener-It\^o chaos expansion
%\end{keywords}

\paragraph{AMS:} 35R60, 60H15, 35K15.
%\begin{AMS}
%\end{AMS}

%\pagestyle{myheadings}
%\thispagestyle{plain}

%%%%%%%%%%%%%%%%%%%%%%
%%% BEGINNING TEXT %%%
%%%%%%%%%%%%%%%%%%%%%%

%%%%%%%%%%%%%%%%%%%%%%%%%%%%%%%%%%%%%%%%%%%%
\section{Introduction}
\label{sec:intro}
%%%%%%%%%%%%%%%%%%%%%%%%%%%%%%%%%%%%%%%%%%%%

We consider the parabolic equation
\begin{equation}
  \label{eq:parabeps}
  \begin{array}{l}
  \pdr{u_\eps}t + P(x,D)u_\eps - \dfrac{1}{\eps^{\frac d2}}
   q\big(\dfrac{x}{\eps}\big) u_\eps =0 \\
   u_\eps(0,x)=u_0(x),
  \end{array} 
\end{equation}
where $P(x,D)$ is an elliptic pseudo-differential operator with
principal symbol of order $\m>d$ and $x\in\Rm^d$.  The initial
condition $u_0(x)$ is assumed to belong to $L^1(\Rm^d)\cap
L^2(\Rm^d)$.  We assume that $q(x)$ is a mean zero, Gaussian,
stationary field defined on a probability space $(\Omega,\mathcal
F,\P)$ with integrable correlation function $R(x)=\E\{q(0)q(x)\}$.

The main objective of this paper is to construct a solution to the
above equation in $L^2(\Omega\times\Rm^d)$ uniformly in time on
bounded intervals (see Theorem \ref{thm:duhameleps} below) and to show
that the solution converges in distribution as $\eps\to0$ to the
unique mild solution of the following stochastic partial differential
equation (SPDE)
\begin{equation}
  \label{eq:SPDE}
  \begin{array}{l}
    \pdr{u}t + P(x,D) u - \sigma u \circ \dot W   =0\\
     u(0,x)=u_0(x),
  \end{array}
\end{equation}
where $\dot W$ denotes spatial white noise, $\circ$ denotes the
Stratonovich product and sigma is defined as
\begin{equation}
  \label{eq:sigma}
  \sigma^2 := (2\pi)^d \hat R(0) = \dint_{\Rm^d} \E \{ q(0)q(x) \} dx.
\end{equation}

We denote by $G(t,x;y)$ the Green's function associated to the above
unperturbed operator. In other words, $G(t,x;y)$ is the distribution
kernel of the operator $e^{-tP(x,D)}$. Our main assumptions on the
unperturbed problem are that $G(t,x;y)=G(t,y;x)$ is continuous and
satisfies the following regularity conditions:
\begin{equation}
  \label{eq:regG}
  \sup_{t,y} \dint_{\Rm^d} |G(t,x;y)| dx
  + \sup_{t,y} \,t^{\frac d\m} \dint_{\Rm^d} |G(t,x;y)|^2 dx 
  +  \sup_{t,x,y} \,t^{\frac{d}{\m}}|G(t,x;y)|\,<\,\infty.
\end{equation}
Note that the $L^2$ bound is a consequence of the $L^1$ and $L^\infty$
bounds.  Such regularity assumptions may be verified e.g. for
parabolic equations with $\m=2$ and $d=1$ or more generally
for equations with $\m=2\n$ an even number and $d<\m$.
The convergence of the random solution to the solution of the SPDE is
obtained under the additional continuity constraint
\begin{equation}
  \label{eq:modulusG}
  \sup_{s\in(0,T),\zeta} s^\gamma \dint_{\Rm^d} |G(s,x,\zeta)-G(s,x+y,\zeta)|dx
   \to 0 \quad \mbox{ as } y\to0\quad \mbox{ for } \quad 
    \gamma=2\Big(1-\dfrac d{\m}\Big).
\end{equation}
Such a constraint may also be verified for Green's functions
of parabolic equations with $\m=2\n$ and $d<\m$; see lemma
\ref{lem:bounds} below.

We look for mild solutions of \eqref{eq:SPDE}, which we recast as
\begin{equation}
  \label{eq:SPDEint}
  u(t,x) = e^{-t P(D)} u_0(x) + \dint_0^t \dint_{\Rm^d}
    G(t-s,x;y) u(s,y) \circ \sigma dW(y) ds.
\end{equation}
Here, $dW$ is the standard Wiener measure on $\Rm^d$ and $\circ$ means
that the integral is defined as a (anticipative) Stratonovich
integral. In section \ref{sec:duhamel}, we define the Stratonovich
integral for an appropriate class of random variables and construct a
solution to the above equation in $L^2(\Omega\times\Rm^d)$ uniformly
in time on bounded intervals by the method of Duhamel expansion; see
Theorem \ref{thm:duhamel} below.  In section \ref{sec:uniq}, we show
that the solution to the above equation is unique in an adapted
functional setting. The convergence of the solution $u_\eps(t)$ to its
limit $u(t)$ is addressed in section \ref{sec:conv}; see Theorem
\ref{thm:conv} below.

\medskip

The analysis of stochastic partial differential equations of the form
\eqref{eq:SPDE} with $\m=2$ and with the Stratonovich product replaced
by an It\^o (Skorohod) product or a Wick product and the white noise
in space replaced by a white noise in space time is well developed; we
refer the reader to e.g.
\cite{D-EJP-99,HOUZ-Birk-96,Ito-SPDE-84,NR-JFA-97,NZ-JFA-89,W-SPDE-86}.
The case of space white noise with It\^o product is analyzed in e.g.
\cite{Hu-PA-02}.  One of the salient features obtained in these
references is that solutions to stochastic equations of the form
\eqref{eq:SPDE} are found to be square-integrable for sufficiently
small spatial dimensions $d$ and to be elements in larger
distributional spaces for larger spatial dimensions; see in particular
\cite{D-EJP-99} for sharp criteria on the existence of locally mean
square random processes solution to stochastic equations. This brings
into question the justification of stochastic models of the form
\eqref{eq:SPDE}.

The theory presented in this paper shows that the solution to
\eqref{eq:SPDE} may indeed be seen as the $\eps\to0$ limit of
solutions to a parabolic equation \eqref{eq:parabeps} with highly
oscillatory coefficient when the spatial dimension is sufficiently
small.  In larger spatial dimensions, the behavior observed in
\cite{B-CMP-1-08} is different. The solution to \eqref{eq:parabeps}
with a properly scaled potential (of amplitude proportional to
$\eps^{-\frac\m2}$ for $\m<d$) converges to the deterministic solution
of a homogenized equation, at least for sufficiently small times. The
solution to a stochastic model no longer represents the asymptotic
behavior of the solution to an equation of the form
\eqref{eq:parabeps} with highly oscillatory random coefficients.

The analysis of equations with highly oscillatory random coefficients
of the form \eqref{eq:parabeps} has also been performed in other
similar contexts. We refer the reader to \cite{PP-GAK-06} for a recent
analysis of the case $\m=2$ and $d=1$ with much more general
potentials than the Gaussian potentials considered in this paper.
When the potential has smaller amplitude, then the limiting solution
as $\eps\to0$ is given by the unperturbed solution of the parabolic
equation where $q$ has been set to $0$. The analysis of the random
fluctuations beyond the unperturbed solution were addressed in e.g.
\cite{B-CLH-08,FOP-SIAP-82}. 

%%%%%%%%%%%%%%%%%%%%%%%%%%%%%%%%%%%%%%%%%%%%
\section{Stratonovich integrals and Duhamel solutions}
\label{sec:duhamel}
%%%%%%%%%%%%%%%%%%%%%%%%%%%%%%%%%%%%%%%%%%%%

The analysis of \eqref{eq:SPDEint} requires that we define the
multi-parameter Stratonovich integral used in the construction of a
solution to the SPDE. The construction of Stratonovich integrals and
their relationships to It\^o integrals is well-studied.  The refer the
reader to e.g.
\cite{DS-SSR-92,HM-SP-88,JK-TAMS-93,NZ-AP-89,SU-SSR-90}.  The
construction that we use below closely follows the functional setting
presented in \cite{J-JTP-06}.  The convergence of processes to
multiple Stratonovich integrals may be found in e.g.
\cite{BJ-SPA-00,BK-SS-97}.

Let $f(x_1,\ldots,x_n)$ be a function of $n$ variables in
$\Rm^d$.  We want to define the iterated Stratonovich integral
$\mI_n(f)$.  Let us first assume that $f$ separates as a product of
$n$ functions defined on $\Rm^d$, i.e.,
$f(x_1,\ldots,x_n)=\prod_{k=1}^n f_k(x_k)$. Then we define
\begin{equation}
  \label{eq:Insimple}
  \mI_n\big(\prod_{k=1}^n f_k(x_k)\big) = \prod_{k=1}^n
   \mI_1(f_k(x_k)),
\end{equation}
where $\mI_1(f)=\int_{\Rm^d} f(x) dW(x)$ is the usual multi-parameter
It\^o integral. It then remains to extend this definition of the integral
to more general functions $f(x)$. 

We define the symmetrized function
\begin{equation}\label{eq:symf}
  f_\ms(x_1,\ldots,x_n) = \dfrac1{n!}\dsum_{{\mathfrak s}\in{\mathfrak S}_n}
    f(x_{\mathfrak s(1)},\ldots,x_{\mathfrak s(n)}),
\end{equation}
where the sum is taken over the $n!$ permutations of the variables
$x_1,\ldots,x_n$. We then define $\mI_n(f)=\mI_n(f_\ms)$ and thus
now consider functions that are symmetric in their arguments.

For the rest of the paper, we write Stratonovich integrals using the
notation $dW$ rather than $\circ \,dW$. For the It\^o convention of
integration, we use the notation $\delta W$. Let $f$ and $g$ be two
functions of $n$ variables. We formally define the inner product
\begin{equation}
  \label{eq:innern}
   \begin{array}{rcl}
  \langle f,g \rangle_n &=& \E\Big\{\dint_{\Rm^{nd}}
   f(x) dW(x_1)\ldots dW(x_n) {\dint_{\Rm^{nd}}
   g(x) dW(x_1)\ldots dW(x_n)} \Big\}\\
   &=& \dint_{\Rm^{2nd}} f(x) {g(x')}
   \E\big\{ dW(x_1)\ldots dW(x_n) {dW(x_{n+1})\ldots dW(x_{2n})}\big\},
   \end{array}
\end{equation}
since the latter has to hold for functions defined as in
\eqref{eq:Insimple}. Here, $x'=(x_{n+1},\ldots,x_{2n})$. We need to
expand the moment of order $2n$ of Gaussian random variables. The moment
is defined as follows:
\begin{equation}
  \label{eq:mtGauss}
  \E\big\{\prod_{k=1}^{2n} dW(x_k)\} = \dsum_{\pp\in\PP}
   \dprod_{k\in A_0(\pp)} \delta(x_k-x_{l(k)}) dx_k dx_{l(k)}.
\end{equation}
Here, $\pp$ runs over all possible pairings of $2n$ variables. There are
\begin{equation}
  \label{eq:cn}
  \rm{card }(\PP) = c_n = \dfrac{(2n-1)!}{(n-1)!2^{n-1}}
    = \dfrac{(2n)!}{n!2^{n}}  = (2n-1)!!
\end{equation}
such pairings. Each pairing is defined by a map $l=l(\pp)$ constructed
as follows. The domain of definition of $l$ is the subset
$A_0=A_0(\pp)$ of $\{1,\ldots,2n\}$ and the image of $l$ is
$B_0=B_0(\pp)=l(A_0)$ defined as the complement of $A_0$ in
$\{1,\ldots,2n\}$. The cardinality of $A_0$ and $B_0$ is thus $n$ and
there are $c_n$ choices of the function $l$ such that $l(k)\geq k+1$.
The formula \eqref{eq:mtGauss} thus generalizes the case $n=1$, where
$\E\{dW(x)dW(y)\}=\delta(x-y)dxdy$.

We extend by density the iterated Stratonovich integral defined in
\eqref{eq:Insimple} to the Banach space $\mB_n$ of functions $f$ that
are bounded for the norm
\begin{equation}
  \label{eq:normIn}
  \|f\|_n = \Big(
   \dsum_{\pp\in\PP} \dint_{\Rm^{2nd}} |f \otimes f|(x_1,\ldots,x_{2n})
   \dprod_{k\in A_0(\pp)} \delta(x_k-x_{l(k)}) dx_k dx_{l(k)}
  \Big)^{\frac12}.
\end{equation}
The above Banach space may be constructed as the completion of smooth
functions with compact support for the above norm \cite{RS-80-I}.
Since the sum of product of functions of one $d-$dimensional variable
are dense in the space of continuous functions, they are dense in the
above Banach space and the Stratonovich integral is thus defined for
such integrands $f(x)$. A more explicit expression may be obtained for
the above norm for functions $f(x)$ that are symmetric in their
arguments.  Since we do not use the explicit expression in this paper,
we shall not derive it explicitly. We note however that
\begin{equation}
  \label{eq:relatIn}
  \|f\|_n^2 = \E\{ \mI_{n+n}(|f\otimes f|) \},
\end{equation}
since $\mI_n(f)\mI_n(f)=\mI_{2n}(f\otimes f)$.

Note that the above space is a Banach subspace of the Hilbert space of
square integrable functions since the $L^2$ norm of $f$ appears for
the pairing $\prod_k\delta(x_k-x_{k+n})$. Note also that the above
space is dense in $L^2(\Rm^{nd})$ for its natural norm. Indeed, let
$f$ be a square integrable function. We can construct a sequence of
functions $f^k$ that vanish on a set of measure $k^{-1}$ in the
vicinity of the sets of measure $0$ where the distributions
$\delta(x_k-x_l)$, $1\leq k,l\leq n$, are supported and equal to $f$
outside of this set. For such functions, we verify that $\|f^k\|_n$ is
the $L^2(\Rm^{nd})$ norm of $f^k$. Moreover, $f^k$ converges to $f$ as
$k\to\infty$ as an application of the dominated Lebesgue convergence
theorem so that $\mB_n$ is dense in $L^2(\Rm^{nd})$. Note finally that
the above expression still defines a norm for functions that are not
necessarily symmetric in their arguments.  This norm applied to
non-symmetric functions is not optimal as far as the definition of
iterated Stratonovich integrals are concerned since many cancellations
may happen by symmetrization \eqref{eq:symf}. However, the above norm
is sufficient in the construction of a Duhamel expansion solution to
the SPDE.

\paragraph{Duhamel solution.} Let us define formally the integral
\begin{equation}
  \label{eq:mHu}
  \mH u(t,x) = \sigma\dint_0^t \dint_{\Rm^d} G(t-s,x;y) u(s,y) dW(y) ds,
\end{equation}
where we recall that $dW$ means an integral in the Stratonovich sense.
The Duhamel solution is defined formally as 
\begin{equation}
  \label{eq:Duhamel}
  u(t,x) = \dsum_{n=0}^\infty u_n(t,x),\quad
  u_{n+1}(t,x) = \mH u_n(t,x),\quad u_0(t,x) = e^{-tP(x,D)}[u_0(x)],
\end{equation}
where $u_0$ is the initial conditions of the stochastic equation,
which we assume is integrable. The above solution is thus defined
formally as a sum of iterated Stratonovich integrals
$u_n(t,x)=\mI_n(f_n(t,x,\cdot))$.

The main result of this section is the following.
\begin{theorem}
  \label{thm:duhamel}
  Let $u(t,x)$ be the function defined in \eqref{eq:Duhamel}.  The
  iterated integrals $u_n(t,x)=\mI_n(f_n(t,x,\cdot))$ are defined in
  $L^2(\Rm^d;\mB_n)$ uniformly in time $t\in (0,T)$ for all $T>0$ and
  $n\geq1$.  When the initial condition $u_0(x)\in L^1(\Rm^d)\cap
  L^2(\Rm^d)$, then $u(t,x)$ is a mild solution to the SPDE in
  $L^2(\Rm^d\times\Omega)$ uniformly in time $t\in (0,T)$ for all
  $T>0$. When $u_0(x)\in L^1(\Rm^d)$, then the deterministic component
  $u_0(t,x)$ in $u(t,x)$ satisfies $t^{\frac d{2\m}}u_0(t,x)\in
  L^2(\Rm^d)$ uniformly in time.
\end{theorem}
\begin{proof}
  The $L^2$ norm of $u(t,x)$ is defined by
  \begin{displaymath}
  \begin{array}{rcl}
  \dint_{\Rm^d} \E\{u^2(t,x)\} dx &=& \dsum_{n,m\geq0} \dint_{\Rm^d}
  \E\{\mI_{n+m} (f_n(t,x,\cdot)\otimes f_m(t,x,\cdot))\} dx\\
  &\leq&\dsum_{n,m\geq0}
    \dint_{\Rm^d}\E\{\mI_{n+m} (|f_n(t,x,\cdot)\otimes f_m(t,x,\cdot)|)\} dx.
   \end{array}
  \end{displaymath}
  We now prove that the latter is bounded uniformly in time on compact
  intervals. The proof shows that $f_n(t,\cdot)$ is also uniformly
  bounded in $L^2(\Rm^d;\mB_n)$ so that the iterated integrals
  $u_n(t,x)$ are indeed well defined. 

  Note that $n+m=2\ban$ for otherwise the above integral vanishes. Then,
  using the notation $t_0=s_0=t$, we have
  \begin{displaymath}
  \begin{array}{l}
  I_{n,m}(t)=\dint_{\Rm^d}\E\{\mI_{n+m}(|f_n(t,x)\otimes f_m(t,x)|)\} dx =\\
  \qquad \dint_{\Rm^d} \dprod_{k=0}^{n-1}\dint_0^{t_k}\dint_{\Rm^{dn}}
   \prod_{k=0}^{n-1} |G|(t_k-t_{k+1},x_k;x_{k+1})
   \Big|\dint_{\Rm^d} G(t_n,x_n;\xi) u_0(\xi) d\xi\Big|
   \dprod_{k=1}^n dt_k\\
   \qquad
   \dprod_{l=0}^{m-1}\dint_0^{s_l}\dint_{\Rm^{dm}}
   \prod_{l=0}^{m-1} |G|(s_l-s_{l+1},y_l;y_{l+1})
   \Big|\dint_{\Rm^d} G(s_m,y_m;\zeta) u_0(\zeta) d\zeta\Big|
   \dprod_{l=1}^m ds_l\\
   \qquad \delta(x_0-x)\delta(y_0-x) \sigma^{n+m}
  \E\{\dprod_{k=1}^n dW(x_k)\dprod_{l=1}^m dW(y_l)\} dx.
  \end{array}
  \end{displaymath}
  Using the fact that $2ab\leq a^2+b^2$ with $a$ and $b$ the Green's
  functions involving $x$ and the fact that $\tau^{\frac d\m} \int
  G^2(\tau,x,y)dx$ is uniformly bounded, we bound the integral in $x$
  by a constant. Let us define $\phi(s)=|t-s|^{-\frac d{\m}}$. As a
  consequence, we obtain that
  \begin{displaymath}
  \begin{array}{l}
  I_{n,m}(t) \!\lesssim\!
   \dint_0^t \!\!\phi(t_1) \dprod_{k=1}^{n-1}\dint_0^{t_k}
    \!\!\!\dint_{\Rm^{d(n-1)}}
   \prod_{k=1}^{n-1} |G|(t_k-t_{k+1},x_k;x_{k+1})
   \Big|\dint_{\Rm^d} \!\!\! G(t_n,x_n;\xi) u_0(\xi) d\xi\Big|
   \dprod_{k=1}^n dt_k\\
   \qquad\quad
   \dint_0^t  \dprod_{l=1}^{m-1}\dint_0^{s_l}\dint_{\Rm^{d(m-1)}}
   \prod_{l=1}^{m-1} |G|(s_l-s_{l+1},y_l;y_{l+1})
   \Big|\dint_{\Rm^d} \!\!\! G(s_m,y_m;\zeta) u_0(\zeta) d\zeta\Big|
   \dprod_{l=1}^m ds_l\\
  \qquad \quad \sigma^{n+m}\E\{\dprod_{k=1}^n dW(x_k)\dprod_{l=1}^m dW(y_l)\}.
  \end{array}
  \end{displaymath}
  Here $a\lesssim b$ means that $a\leq Cb$ for some constant $C>0$.
  
  Let us re-label $x_{n+l}=y_l$ and $t_{n+l}=s_l$ for $1\leq l\leq
  m$. We also define $\bx=(x_0,\ldots,x_{n+m+1})$. Then we find that
  \begin{align*}
     I_{n,m}(t) &\leq \sigma^{n+m}
   \dint_{\Rm^{2\ban d}}   H_{n,m}(t,\bx)\,
    \E\Big\{\dprod_{k=1}^{2\ban} dW(x_k)\Big\},\\
  H_{n,m}(t,\bx)&= \dint_0^t \phi(t_1)\dprod_{k=1}^{n-1}\dint_0^{t_k}
   \dint_0^t \dprod_{l=1}^{m-1}\dint_0^{t_{n+1+l}}
  \dprod_{k=1, k\not=n}^{n+m-1}|G|(t_k-t_{k+1},x_k;x_{k+1})
  \\\qquad&\Big|\dint_{\Rm^{d}} G(t_n,x_n;\xi) u_0(\xi) d\xi\Big|
   \Big|\dint_{\Rm^{d}} G(t_{n+m},x_{n+m};\zeta) u_0(\zeta) d\zeta\Big|
   \dprod_{k=1}^{n+m} dt_k.
  \end{align*}
  We now recall the pairings introduced in \eqref{eq:mtGauss} and
  replace $n$ by $\ban$ there.  Let us introduce the notation
  \begin{displaymath}
  y_k = \left\{
    \begin{array}{ll}
       x_{k+1} & k\not=n,n+m \\
       \xi & k=n \\
       \zeta & k=n+m
    \end{array}\right. \qquad
   \tau_k = \left\{
    \begin{array}{ll}
       t_{k+1} & k\not=n,n+m \\
       0 & k=n,n+m,
    \end{array}\right.
  \end{displaymath}
  so that $H_{n,m}(t,\bx)$ is bounded by
  \begin{displaymath}
     \dint_{\Rm^{2d}} \dint_0^t\!\phi(t_1)
   \dprod_{k=1}^{n-1}\dint_0^{t_k}
   \dint_0^t\dprod_{l=1}^{m-1}\dint_0^{t_{n+l}}
   \dprod_{k=1}^{2\ban} |G|(t_k-\tau_k,x_k;y_k)
    |u_0(\xi)|\,|u_0(\zeta)|\, d\xi d\zeta
    \!\! \dprod_{k=1}^{n+m}\!\! dt_k.
  \end{displaymath}
  Now, we have for each pairing $\pp\in\PP$,
  \begin{displaymath}
    \dprod_{k=1}^{2\ban} |G|(t_k-\tau_k,x_k;y_k) = 
   \dprod_{k\in A_0}  |G|(t_k-\tau_k,x_k;y_k) 
    |G|(t_{l(k)}-\tau_{l(k)},x_{l(k)};y_{l(k)}),
  \end{displaymath}
  and as a consequence, using the delta functions appearing in
  \eqref{eq:mtGauss},
  \begin{equation}\label{eq:Imn1}
  \begin{array}{l}
  I_{n,m}(t) \leq \sigma^{n+m}\dsum_{\pp\in\PP}\dint_{\Rm^{2d}}
   \dint_0^t\!\phi(t_1)\dprod_{k=1}^{n-1}\dint_0^{t_k}
   \dint_0^t\dprod_{l=1}^{m-1}\dint_0^{t_{n+l}}
    |u_0(\xi)|\,|u_0(\zeta)|\\
   \qquad \dint_{\Rm^{\ban d}}\dprod_{k\in A_0} \Big(  |G|(t_k-\tau_k,x_k;y_k) 
    |G|(t_{l(k)}-\tau_{l(k)},x_{k};y_{l(k)}) dx_k \Big)
    \, d\xi d\zeta
    \!\! \dprod_{k=1}^{n+m}\!\! dt_k\\
   \qquad \leq \dsum_{\pp\in\PP} \dint_0^t\!\phi(t_1)
   \dprod_{k=1}^{n-1}\dint_0^{t_k}
   \dint_0^t\dprod_{l=1}^{m-1}\dint_0^{t_{n+l}}\dprod_{k\in A_0}
    \dfrac{C}{(t_{\mk(k)}-\tau_{\mk(k)})^{\frac d\m}} \, \|u_0\|^2_{L^1}
   \!\! \dprod_{k=1}^{n+m}\!\! dt_k,
  \end{array}
  \end{equation}
  for some positive constant $C$ in which we absorb $\sigma^2$.  On
  the second line above, the $y_{l(k)}$ are evaluated at
  $x_{l(k)}=x_k$.  The function $k\mapsto \mk(k)$ for $k\in A_0$ is at
  the moment an arbitrary function such that $\mk(k)=k$ or
  $\mk(k)=l(k)$.  The last line is obtained iteratively in increasing
  values of $k$ in $A_0$ by using that one of the Green's function is
  integrable in $x_k$ uniformly in the other variables and that the
  other Green's function is bounded independent of the spatial
  variables by a constant times the time variable to the power
  $-\alpha$ with $\alpha:=\frac d\m$.  We have used here assumption
  \eqref{eq:regG}. It then remains to integrate in the variables $\xi$
  and $\zeta$ and we use the initial condition $u_0(x)$ for this.
  
  Let us now choose the map $\mk(k)$. It is constructed as follows.
  When both $k$ and $l(k)$ belong to $\{1,\ldots,n\}$ or both belong
  to $\{n+1,n+m\}$, then we set $\mk(k)=k$. When $k\in\{1,\ldots,n\}$
  and $l(k)\in\{n+1,n+m\}$ (i.e., when there is a crossing from the
  $n$ first variables to the $m$ last variables), then we choose
  $\mk(k)=k$ for half of these crossings and $\mk(k)=l(k)$ for the
  other half. When the number of crossings is odd, the last crossing
  is chosen with $\mk(k)=k$. 
  
  Let us define $A_0^1=\mk(A_0)\cap\{0,\ldots,n\}$ and
  $A_0^2=\mk(A_0)\backslash A_0^1$.  Let $n_0=n_0(\pp)$ be the number
  of elements in $A_0^1$ and $m_0=m_0(\pp)$ be the number of elements
  in $A_0^2$ such that $n_0+m_0=\ban$. Let $p=p(\pp)$ be the number of
  crossings in $\pp$.  Then, by construction of $m$, we have
  \begin{equation}\label{eq:n0m0}
  n_0 = \dfrac{n-p}2 + \Big[\dfrac{p+1}2\Big],\qquad
  m_0 = \dfrac{m-p}2 + \Big[\dfrac{p}2\Big],
  \end{equation}
  where $[\frac{p+1}2]=\frac{p+1}2$ if $p$ is odd and $\frac p2$ if
  $p$ is even, with $[\frac{p+1}2]+[\frac p2]=p$. Thus, $n_0$ is
  bounded by $\frac{n+1}2$ and $m_0$ by $\frac{m}2$.

  We thus obtain that
  \begin{align*}
  I_{n,m}(t) \leq C^{\ban} \|u_0\|^2_{L^1}  \sum_{\pp\in\PP}\,\,
  &\Big[ \prod_{k=0}^{n-1} \int_0^{t_k} \!\phi(t_1)\dprod_{k\in A_0^1}
    \dfrac1{(t_k-t_{k+1})^\alpha} \dprod_{k=1}^n dt_k \Big]\\
  &\Big[ \dprod_{l=0}^{m-1} \dint_0^{s_l} \dprod_{n+l\in A_0^2}
    \dfrac1{(s_l-s_{l+1})^\alpha} \dprod_{l=1}^m ds_l \Big],
  \end{align*}
  with the convention that $t_0=s_0=t$, $t_{n+1}=0$ and $s_{m+1}=0$.
  It remains to estimate the time integrals, which are very small, and
  sum over a very large number of them. It turns out that these
  integrals admit explicit expressions. The construction of the
  mapping $\mk(k)$ ensures that the number of singular terms of the
  form $\tau^{-\alpha}$ is not too large in the integrals over the $t$
  and the $s$ variables.
  
  Let $\alpha_k$ for $0\leq k\leq n$ be defined such that
  $\alpha_0=\alpha$, $\alpha_k=\alpha$ for $k\in A_0^1$ and
  $\alpha_k=0$ otherwise.  Still with the convention that $t_{n+1}=0$,
  we thus want to estimate
  \begin{equation} \label{eq:In}
  I_n=I_n(\pp) = \prod_{k=0}^{n-1} \int_0^{t_k} \dprod_{k=0}^n
    \dfrac1{(t_k-t_{k+1})^{\alpha_k}} \dprod_{k=1}^n dt_k.
  \end{equation}
  The integrals are calculated as follows. Let us consider the last
  integral:
  \begin{displaymath}
  \dint_0^{t_{n-1}} \dfrac{1}{(t_{n-1}-t_n)^{\alpha_{n-1}}}
   \dfrac{1}{t_n^{\alpha_n}} dt_n
   = t_{n-1}^{1-\beta_{n-1}} \dint_0^1 
   \dfrac{1}{(1-u)^{\alpha_{n-1}}u^{\alpha_n}}  du,
  \end{displaymath}
  where we define $\beta_n=\alpha_n$ and
  $\beta_m=\beta_{m+1}+\alpha_m$ for $0\leq m\leq n-1$. The latter
  integral is thus given by
  \begin{displaymath}
  t_{n-1}^{1-\beta_{n-1}} B(1-\beta_n,1-\alpha_{n-1}),
  \end{displaymath}
  where $B(x,y)=\frac{\Gamma(x)\Gamma(y)}{\Gamma(x+y)}$ is the Beta
  function and $\Gamma(x)$ the Gamma function equal to $(x-1)!$ for
  $x\in\Nm^*$. The integration in $t_{n-2}$ then yields
  \begin{displaymath}
  \dint_0^{t_{n-2}} \dfrac{ t_{n-1}^{1-\beta_{n-1}} }
  {(t_{n-2}-t_{n-1})^{\alpha_{n-2}}} dt_{n-2}
   = t_{n-2}^{2-\beta_{n-2}} B(2-\beta_{n-1},1-\alpha_{n-2}).
  \end{displaymath}
  By induction, we thus obtain that
  \begin{equation}\label{eq:solIn}
  I_n = t_0^{n-\beta_0} \prod_{k=0}^{n-1}
    B(n-k-\beta_{k+1},1-\alpha_k)
  = t_0^{n-\beta_0} \prod_{k=0}^{n-1} 
   \dfrac{\Gamma(n-k-\beta_{k+1})\Gamma(1-\alpha_k)}
    {\Gamma(n-k+1-\beta_{k+1}-\alpha_k)}.
  \end{equation}
  Since $\beta_{k+1}+\alpha_k=\beta_k$, we obtain by telescopic
  cancellations that
  \begin{displaymath}
  I_n = t_0^{n-\beta_0}\dfrac{\Gamma(1-\beta_n)}{\Gamma(n+1-\beta_0)}
   \prod_{k=0}^{n-1} \Gamma(1-\alpha_k).
  \end{displaymath}
  Then with our explicit choices for the coefficients $\alpha_k$
  above, we find that $\beta_0=(n_0+1)\alpha$ so that
  \begin{displaymath}
  I_n = t_0^{n-(n_0+1)\alpha}\dfrac{\Gamma(1-\alpha_n) \Gamma^{n_0}(1-\alpha)}
   {\Gamma(n+1-(n_0+1)\alpha)}.
  \end{displaymath}
  
  For a fixed $\pp$, we see that the contribution of the time
  integrals in $I_{n,m}(t)$ is bounded by a constant (since
  $\Gamma(1-\alpha)$ is bounded as $\alpha<1$) times
  \begin{displaymath}
  \dfrac{\Gamma^{\ban}(1-\alpha)}
   {\Gamma(n+1-(n_0+1)\alpha)\Gamma(m+1-m_0\alpha)}
 \leq \dfrac{\Gamma^{\ban}(1-\alpha)}
   {\Gamma((n+1)(1-\frac{\alpha}2)-\alpha)\Gamma((m+1)(1-\frac\alpha2))}
  \end{displaymath}
  based on the values of $n_0$ and $m_0$. Using Stirling's formula
  $\Gamma(z)\sim (\frac{2\pi}z)^{\frac12}(\frac ze)^z$ so that
  $\Gamma(z)$ is bounded from below by $(\frac{z}C)^z$ for $C<e$, we
  find that the latter term is bounded by
  \begin{equation} \label{eq:bdnm}
     \dfrac{C^{n+m}}{n^{n(1-\frac\alpha2)}m^{m(1-\frac\alpha2)}},
  \end{equation}
  for some positive constant $C$. The latter bound holds for each
  $\pp\in\PP$.  Using the Stirling formula again, we observe that the
  number of graphs in $\PP$ is bounded by $(\frac{2\ban}e)^\ban$. As a
  consequence, we have
  \begin{equation}\label{eq:finalImn}
     I_{m,n} \leq t_0^{n+m-\alpha\ban}\Big(\dfrac{2\ban}e\Big)^\ban 
    \dfrac{C^n C^m}{n^{n(1-\frac{\alpha}2)}m^{m(1-\frac{\alpha}2)}}
  \end{equation}
  Using the concavity of the log function, we have
  \begin{displaymath}
  n^n m^m \geq \Big(\dfrac{n^2+m^2}{n+m}\Big)^{n+m}
   \geq \Big(\dfrac{n+m}{2}\Big)^{n+m},
  \end{displaymath}
  so that
  \begin{displaymath}
  \ban^\ban \leq C^nC^m n^{\frac n2} m^{\frac m2}.
  \end{displaymath}
  As a consequence, we have
  \begin{equation} \label{eq:Jmn}
    I_{n,m}  \leq   J_{n,m}(t) := t_0^{(n+m)(1-\frac\alpha2)-\alpha}  C^n C^m 
   \dfrac{1}{n^{\frac n2(1-\alpha)}m^{\frac m2(1-\alpha)}}.
  \end{equation}
  The bound with $n=m$ shows that for $n\geq1$, $u_n(t,x)$ belongs to
  $L^2(\Rm^d;\mB_n)$ uniformly in time on compact intervals since
  $2(1-\alpha)>0$.  Now the deterministic component $u_0(t,x)$ is in
  $L^2(\Rm^d)$ uniformly in time when $u_0(x)\in L^2(\Rm^d)$ while
  $t^{\frac\alpha2}u_0(t,x)$ is in $L^2(\Rm^d)$ uniformly in time when
  $u_0(x)\in L^1(\Rm^d)$.  Upon summing the above bound over $n$ and
  $m$, we indeed deduce that $u(t,x)$ belongs to
  $L^2(\Omega\times\Rm^d)$ uniformly in time on compact intervals when
  $u_0\in L^2(\Rm^d)$.
  
  The above uniform convergence shows that $\mH u(t,x)$ is well
  defined in $L^2(\Omega\times\Rm^d)$ uniformly in time. Moreover, we
  verify that $\mH u(t,x)=\sum_{n\geq1}u_n(t,x) = u(t,x)-u_0(t,x)$. 
  This shows that $u(t,x)$ is a mild solution of the stochastic partial
  differential equation and concludes the proof of the theorem.
\end{proof}
%%%%%%%%%%%%%%%%%%%%%%%%%%%%%%%%%%%%%%%%%%%%
\section{Uniqueness of the SPDE solution}
\label{sec:uniq}
%%%%%%%%%%%%%%%%%%%%%%%%%%%%%%%%%%%%%%%%%%%%

Let us assume that two solutions exist in a linear vector space $\mM$.
Then their difference, which we call $u$, solves the equation
\begin{displaymath}
  u = \mH u = \mH^p u,
\end{displaymath}
for all $p\geq0$. The space $\mM$ is defined so that $\mH^p u$ is
well-defined and is constructed as follows.

We construct $u\in \mM$ as a sum of iterated Stratonovich integrals
\begin{displaymath}
  u(t,x) = \dsum _{m\geq0} \mI_n(f_n(t,x,\cdot)).
\end{displaymath}
Because the iterated Stratonovich integrals do not form an orthogonal
basis of random variables in $L^2(\Omega)$, the above sum is formal 
and needs to be defined carefully. We need to ensure that the 
sum converges in an appropriate sense and that $\mM$ is closed
under the application of $\mH$.

One way to do so is to construct $u(t,x)$ using the classical
Wiener-It\^o chaos expansion
\begin{displaymath}
  u(t,x) = \dsum_{m\geq0} I_m(g_m(t,x,\cdot)),
\end{displaymath}
where $I_m$ is the iterated It\^o integral, and to show that 
the above series is well defined. We then also impose that
the chaos expansion of $\mH^pu$ is also well-defined. 

We first need a calculus to change variables from a definition in
terms of iterated Stratonovich integrals to one in terms of iterated
It\^o integrals. This is done by using the Hu-Meyer formulas. We
re-derive this expression as follows.  We denote by $\delta W$ an
It\^o integral and by $dW$ a Stratonovich integral. We project
Stratonovich integrals onto the orthogonal basis of It\^o integrals as
follows
\begin{displaymath}
  \E\{\mI_n(f_n) I_m(\phi_m)\} = 
  \E\{I_m(g_m)I_m(\phi_m)\} = m! \dint_{\Rm^{md}} g_m \phi_m dx,
\end{displaymath}
where $\phi_m$ is a test function. We find that 
$\E\{\mI_n(f_n) I_m(\phi_m)\}$ is equal to
\begin{displaymath}
  \dint_{\Rm^{(n+m)d}}
  f_n(x_1,\ldots,x_n) \phi_m(y_1,\ldots,y_m)
   \E\{dW(x_1)\ldots dW(x_n) \delta W(y_1)\ldots \delta W(y_m)\}.
\end{displaymath}
The moment of product of Gaussian variables is handled as in
\eqref{eq:mtGauss} with the exception that $\E\{\delta W(y_k)\delta
W(y_l)\}=0$ for $k\not=l$ by renormalization of the It\^o-Skorohod
integral. The functions $f_n$ and $\phi_m$ are symmetric in their
arguments (i.e., invariant by permutation of its variables). We
observe that the variables $y$ need be paired with $m$ variables $x$.
There are ${n\choose m}$ ways of pairing the $y$ variables. There
remain $n-m=2k$ variables that need be paired, for a possible number
of pairings equal to
\begin{displaymath}
  \dfrac{(2k-1)!}{(k-1)!2^{k-1}}.
\end{displaymath}
The above term is thus given by
\begin{displaymath}
  {m+2k \choose m}\dfrac{(2k-1)!}{(k-1)!2^{k-1}}
  \dint \Big( \dint f_{m+2k}(y_1,\ldots y_m, x_1,x_1,\ldots x_k,x_k)
   \prod_{l=1}^k dx_l \Big) \phi_m(y_1,\ldots y_m) \prod_{p=1}^m dy_p.
\end{displaymath}
This shows that $g_m$ is given by 
\begin{displaymath}
  g_m(x_1,\ldots, x_m) = \dfrac{(m+2k)!}{m! k! 2^k}
   \dint f_{m+2k} (x_1,\ldots, x_m, y_1^{\otimes 2},\ldots,
    y_k^{\otimes 2}) \prod_{p=1}^k dy_k.
\end{displaymath}
Here $y^{\otimes 2}\equiv (y,y)$. The coefficients $g_m$ are therefore
obtained by integrating $n-m$ factors pairwise in the coefficients
$f_n$.  This allows us to write the iterated Stratonovich integral as
a sum of iterated It\^o integrals as follows:
\begin{displaymath}
  \mI_n(f_n) = \dsum_{k=0}^{[\frac n2]}
  \dfrac{n!}{(n-2k)!k!2^k} I_{n-2k}\big(\dint_{\Rm^{kd}} 
    f_n(x_{n-2k},y^{\otimes 2})dy\big).
\end{displaymath}
This is the Hu-Meyer formula. More interesting for us is the reverse
change of coordinates. Let us define formally
\begin{displaymath}
  f = \dsum_{n\geq0} \mI_n(f_n) = \dsum_{m\geq0} I_m(g_m).
\end{displaymath}
Then we find that
\begin{displaymath}
  g_m(x) = \dsum_{k\geq0} \dfrac{(m+2k)!}{m!k!2^k}
 \dint_{\Rm^{kd}} f_{m+2k}(x,y^{\otimes 2})dy.
\end{displaymath}
The square integrability of the coefficients $g_m$ is a necessary
condition for the random variables $f$ to be square integrable, and
more generally, to be in larger spaces of distributions
\cite{HOUZ-Birk-96}. The above formula provide the type of constraints
we need to impose on the traces of the coefficients $f_n$. For square
integrable variables, we consider the normed vector space $\mM_f$ of
random variables
\begin{displaymath}
  f = \dsum_{n\geq0} \mI_n(f_n)
\end{displaymath}
where the coefficients $\{f_n\}$ are bounded for the norm
\begin{displaymath}
  \|f\|_{\mM_f} = \Big(\dsum_{m\geq0} m! \dint \Big(\dsum_{k\geq0} 
  \dfrac{(m+2k)!}{m!k!2^k}
 \dint_{\Rm^{kd}} |f_{m+2k}|(x,y^{\otimes 2})dy\Big)^2 dx\Big)^{\frac12} 
  <\infty.
\end{displaymath}
Note that the above defines a norm as the triangle inequality is
clearly satisfied and for $k=0$, we find that the $L^2$ norm of each
$f_m$ has to vanish, so that $f_m\equiv0$ for all $m$ when the norm
vanishes. Note also that $\mM_f$ is a dense subset of $L^2(\Omega)$ as
any square integrable function $g_m$ may be approximated by a function
$f_m^k$, which vanishes in a set of Lebesgue measure at most $k^{-1}$
in the vicinity of the measure $0$ set of diagonals given by the
support of the distributions $\delta(x_k-x_l)$. For such functions, we
verify that $f_m^k=g_m^k$ so that the It\^o and Stratonovich iterated
integrals agree. We also have that $g_m^k$ converges to $g_m$ by
density. Since every square integrable random variable may be
approximated by a finite number of terms in the chaos expansion, this
concludes our proof that $\mM_f$ is dense in $L^2(\Omega)$ equipped with
its natural metric.

Let us now move to the analysis of the stochastic integral $\mH$.  It
turns out that $\mM_f$ is not stable under $\mH$ nor is it in any
natural generalization of $\mM_f$. Let us define
\begin{displaymath}
  u(t,x) = \dsum_{n\geq 0} \mI_n(f_n(t,x,\cdot)),\qquad
  \mH u(t,x) = \dsum_{n\geq 0} \mI_n((\mH f)_n(t,x,\cdot))
\end{displaymath}
We then observe that 
\begin{displaymath}
  \mH f_{n+1}(t,x,y) = \sigma 
   \ms\big[\dint_0^t G(t-s,x,y_1) f_n(s,y) ds\big],
\end{displaymath}
where $\ms$ is the symmetrization with respect to the
$d(n+1)-$dimensional $y$ variables.  Let us consider $\mH^2 f_{n+2}$,
which depends only on $f_n$.  Let $c_{m,k}=\frac{(m+2k)!}{m!k!2^k}$
the coefficient that appears in the definition of $g_m$. Then, for
$\mH^2 f_{n+2}$ relative to $f_n$, the coefficients indexed by $k$ are
essentially replaced by coefficients indexed by $k+1$. Since
$c_{m,k+1}$ is not bounded by a multiple of $c_{m,k}$ uniformly, the
integral operator $\mH^2$ cannot be bounded in $\mM_f$. The reason why
solutions to the stochastic equation may still be found is because the
integrations in time after $n$ iterations of the integral $\mH$
provide a factor inversely proportional to $n!$. This factors allows
us to stabilize the growth in the traces that appears by going from
$c_{m,k}$ to $c_{m,k+1}$.  Uniqueness of the solution may thus only be
obtained in a space where the factor $n!$ appears, at least
implicitly.

A suitable functional space is constructed as follows. Let $g_m$ be
the chaos expansion coefficients associated to the coefficients
$|f_n|$ and $g_{m,p}$ the chaos expansion coefficients associated to
the coefficients $|\mH^p f_n|$.

Then we impose that the coefficients $\{f_n\}$ be bounded for the norm
\begin{equation} \label{eq:defM}
  \sup_{m\geq0} \,\,\sup_{p\geq0} \,\,\sup_{t\in(0,T)}\,
  \Big(c_p \dint g^2_{m,p}(t,x,y) dx dy\Big)^{\frac12} < \infty,
\end{equation}
where $c_p$ is an increasing series such that $c_p\to\infty$ as
$p\to\infty$. Here $T$ is a fixed (arbitrary) positive time. We denote
by $\mM=\mM(T)$ the normed vector space of random fields $u(t,x)$ for
which the decomposition in iterated Stratonovich integrals satisfies
the above constraint.

We are now ready to state the main result of this section.
\begin{theorem}
  \label{thm:uniq}
  Let $T>0$ be an arbitrary time and $u_0(x)\in L^1(\Rm^d)\cap
  L^2(\Rm^d)$. The solution constructed in Theorem \ref{thm:duhamel} is
  the unique mild solution to the stochastic partial differential
  equation \eqref{eq:SPDE} in the space $\mM=\mM(T)$.
\end{theorem}

\begin{proof}
  Let us first prove uniqueness in $\mM$. We have $u=\mH^p u$ for all
  $p\geq0$.  This implies that $g_m(t,x,\cdot)=g_{m,p}(t,x\cdot)$. The
  latter converges to $0$ in the $L^2$ sense as $p\to\infty$. This
  implies that $g_m(t,x,\cdot)$ uniformly vanishes for all $m$ so that
  $u\equiv0$.
  
  Let now $u(t,x)$ be given by the following Duhamel expansion
  \begin{displaymath}
  u(t,x) = \dsum_{n\geq0} u_n(t,x),\quad
  u_{n+1}(t,x) = \mH u_n (t,x) = \mH^{n+1} u_0(t,x),\quad
  u_0(t,x) = e^{-tP} u_0(x). 
  \end{displaymath}
  We thus verify that
  \begin{displaymath}
  \mH^k u(t,x) = \dsum_{n\geq k} u_n(t,x).
  \end{displaymath}
  This shows that the $L^2$ norm of $\mH^k u(t,\cdot)$ is bounded by
  the sum of the coefficients $I_{n,m}(t)$ for $n,m\geq k$. This sum
  clearly converges to $0$ as $k\to\infty$. Call this sum $c_k^{-1}$.
  We recall that
  \begin{displaymath}
  \dint \E\{ u^2(t,x)\} dx = \dsum_{m\geq0} m! \dint g_m^2(t,x,y)dxdy
   = \E \dint \dsum_{m,n} \mI_{n+m}(f_n\otimes f_m)(t,x)dx.
  \end{displaymath}
  Then we find that
  \begin{displaymath}
  c_k \dint g^2_{m,k}(t,x,y) dx dy \leq c_k 
   \|u(t,\cdot)\|_{L^2(\Omega\times\Rm^d)}^2 \leq C.
  \end{displaymath}
  This shows that $u$ belongs to $\mM$.
\end{proof}
The same theory holds when the supremum in $m$ is replaced by a sum
with weight $m!$ so that $\mM$ becomes a subspace of $L^2$. In some
sense, the subspace created above is the smallest we can consider that
is stable under application of $\mH$. When $u_0(x)\in L^1(\Rm^d)$ not
necessarily in $L^2(\Rm^d)$, then the deterministic component
$u_0(t,x)$ is not square integrable uniformly in time. The space 
$\mM$ may then be replaced by a different space where 
$c_p$ in \eqref{eq:defM} is replaced by $t^{\frac\alpha2}c_p$.
%%%%%%%%%%%%%%%%%%%%%%%%%%%%%%%%%%%%%%%%%%%%
\section{Convergence result}
\label{sec:conv}
%%%%%%%%%%%%%%%%%%%%%%%%%%%%%%%%%%%%%%%%%%%%

Let us now come back to the solution of the equation with random
coefficients \eqref{eq:parabeps}. The theory of existence for such
an equation is very similar to that for the stochastic limit. We define
formally the integral
\begin{equation}
  \label{eq:mHu}
  \mH_\eps u(t,x) = \dint_0^t \dint_{\Rm^d} G(t-s,x;y) u(s,y) 
   q_\eps(y) dy ds,
\end{equation}
where we have defined $q_\eps(y)=\eps^{-\frac d2}q(\frac y\eps)$.
The Duhamel solution is defined formally as 
\begin{equation}
  \label{eq:Duhameleps}
  u_\eps(t,x) = \dsum_{n=0}^\infty u_{n,\eps}(t,x),\,\,\,
  u_{n+1,\eps}(t,x) = \mH_\eps u_{n,\eps}(t,x),\,\,\, u_0(t,x) 
 = e^{-tP(x,D)}[u_0(x)],
\end{equation}
where $u_0$ is the initial conditions of the stochastic equation,
which we assume is integrable.  We have the first result:
\begin{theorem}
  \label{thm:duhameleps}
  The function $u_\eps(t,x)$ defined in \eqref{eq:Duhameleps} solves
  \begin{equation}
    \label{eq:mildeps}
    u_\eps(t,x)=\mH_\eps u_\eps(t,x) + e^{-tP(x,D)}[u_0(x)],
  \end{equation}
  and is in $L^2(\Rm^d\times\Omega)$ uniformly in time $t\in (0,T)$
  for all $T>0$.
\end{theorem}
\begin{proof}
  The proof goes along the same lines as that of Theorem \ref{thm:duhamel}.
   The $L^2$ norm of $u_\eps(t,x)$ is defined by
  \begin{displaymath}
  \begin{array}{rcl}
  \dint_{\Rm^d} \E\{u_\eps^2(t,x)\} dx &=& \dsum_{n,m\geq0} \dint_{\Rm^d}
  \E\{u_{n,\eps}(t,x)u_{m,\eps}(t,x)\} dx\\
  &\leq&\dsum_{n,m\geq0}
    \dint_{\Rm^d}\E\{|u_{n,\eps}|(t,x)|u_{m,\eps}|(t,x)\} dx 
   \leq I_{m,n,\eps}(t),
   \end{array}
  \end{displaymath}
  where
  \begin{displaymath}
  \begin{array}{l}
  I_{n,m,\eps}(t)
    = \dint_{\Rm^d} \dprod_{k=0}^{n-1}\dint_0^{t_k}\dint_{\Rm^{dn}}
   \prod_{k=0}^{n-1} |G|(t_k-t_{k+1},x_k;x_{k+1})
   \Big|\dint_{\Rm^d} G(t_n,x_n;\xi) u_0(\xi) d\xi\Big|
   \dprod_{k=1}^n dt_k\\
   \qquad
   \dprod_{l=0}^{m-1}\dint_0^{s_l}\dint_{\Rm^{dm}}
   \prod_{l=0}^{m-1} |G|(s_l-s_{l+1},y_l;y_{l+1})
   \Big|\dint_{\Rm^d} G(s_m,y_m;\zeta) u_0(\zeta) d\zeta\Big|
   \dprod_{l=1}^m ds_l\\
   \qquad \delta(x_0-x)\delta(y_0-x)
  \E\{\dprod_{k=1}^n q_\eps(x_k)dx_k\dprod_{l=1}^m q_\eps(y_l)dy_l\} dx.
  \end{array}
  \end{displaymath}
  Following the proof of Theorem \ref{thm:duhamel}, we obtain
  \begin{displaymath}
    I_{n,m,\eps}(t) \leq \dint_{\Rm^{2\ban d}}   H_{n,m}(t,\bx)\,
     \E\Big\{\dprod_{k=1}^{2\ban} q_\eps(x_k)dx_k \Big\}.
  \end{displaymath}
  The statement \eqref{eq:mtGauss} now becomes
  \begin{equation}
    \label{eq:mtGausseps}
    \E\Big\{\dprod_{k=1}^{2\ban} q_\eps(x_k)dx_k \Big\}
   =  \sum_{\pp\in\PP}\dprod_{k\in A_0(\pp)} 
     \eps^{-d}R\big(\frac{x_k-x_{l(k)}}\eps\big) dx_k dx_{l(k)},
  \end{equation}
  where we recall that $R(x)=\E\{q(0)q(x)\}$ is the correlation
  function of the Gaussian field $q$.  This yields
   \begin{displaymath}
  \begin{array}{l}
  I_{n,m,\eps}(t) \leq \dsum_{\pp\in\PP}\dint_{\Rm^{2d}}
   \dint_0^t\phi(t_1)\dprod_{k=1}^{n-1}\dint_0^{t_k}
   \dint_0^t\dprod_{l=1}^{m-1}\dint_0^{t_{n+l}}
    |u_0(\xi)|\,|u_0(\zeta)| \dint_{\Rm^{\ban d}}\dprod_{k\in A_0} \\
   \,\,\, \Big(  |G|(t_k-\tau_k,x_k;y_k) 
    |G|(t_{l(k)}-\tau_{l(k)},x_{l(k)};y_{l(k)}) 
   \eps^{-d}\Big|R\big(\dfrac{x_k-x_{l(k)}}\eps\big)\Big|
    dx_k dx_{l(k)} \Big)
    \, d\xi d\zeta
    \!\! \dprod_{k=1}^{n+m}\!\! dt_k.
  \end{array}
  \end{displaymath}
  For each $k\in A_0$ considered iteratively with increasing order,
  the term between parentheses is bounded by the $L^1$ norm of the
  Green's function integrated in $x_{\mk(k)}$ times the integral of
  the correlation function in the variable $x_{\mk'(k)}$, with
  $(\mk(k),\mk'(k))=(k,l(k))$, which gives a $\sigma^2$ contribution
  thanks to the definition \eqref{eq:sigma}, times the $L^\infty$ norm
  of the Green's function in the variable $x_{\mk'(k)}$. Using
  \eqref{eq:regG} and the integrability of the correlation function
  $R(x)$, this shows that
  \begin{displaymath}
    I_{n,m,\eps}(t) \leq 
   \dsum_{\pp\in\PP} \dint_0^t\phi(t_1)\dprod_{k=1}^{n-1}\dint_0^{t_k}
   \dint_0^t\dprod_{l=1}^{m-1}\dint_0^{t_{n+l}}\dprod_{k\in A_0}
    \dfrac{C}{(t_{\mk(k)}-\tau_{\mk(k)})^{\frac d\m}} \, \|u_0\|^2_{L^1}
   \!\! \dprod_{k=1}^{n+m}\!\! dt_k,
  \end{displaymath}
  as in the proof of Theorem \ref{thm:duhamel}. The rest of the proof is
  therefore as in Theorem \ref{thm:duhamel} and shows that each
  $u_{n,\eps}(t,x)$ is well defined in $L^2(\Rm^d\times\Omega)$
  uniformly in time and that the series defining $u(t,x)$ converges
  uniformly in the same sense.
\end{proof}

\paragraph{Mollification and convergence result.}
We now have defined a sequence of solutions $u_\eps(t,x)$ and a
limiting solution $u(t,x)$. When $q_\eps$ and the white noise $W$ used
in the construction of $u(t,x)$ are independent, then the best we can
hope for is that $u_\eps$ converges in distribution to $u$. The
convergence is in fact much stronger (path-wise) by constructing
$q_\eps dx$ as a mollifier of $dW$. Let $\hat R(\xi)$ be the 
power spectrum of $q$, which is defined as the Fourier transform of 
$R(x)$. By Bochner's theorem, the power spectrum is non-negative and
we may define $\hat \rho(\xi)=(\hat R(\xi))^{\frac12}$. Let $\rho(x)$
be the inverse Fourier transform of $\hat \rho$. We may then define
\begin{equation}\label{eq:tildeq}
  \tilde q(x) = \dint_{\Rm^d} \rho(x-y) dW(y),
\end{equation}
and obtain a stationary Gaussian process $\tilde q(x)$. This process
is mean-zero and its correlation function is given by
\begin{displaymath}
  \tilde R(x) = \dint_{\Rm^d} \rho(x-y) \rho(y) dy = R(x),
\end{displaymath}
by inverse Fourier transform of a product. As a consequence, $q(x)$
and $\tilde q(x)$ have the same law since they are mean zero and their
correlation functions agree. The corresponding Duhamel solutions
$u_\eps$ and $\tilde u_\eps$ also have the same law by inspection.  It
thus obviously remains to understand the limiting law of $\tilde
u_\eps$ to obtain that of $u_\eps$. It turns out that $\tilde u_\eps$
may be interpreted as a mollifier of $u(t,x)$, the solution
constructed in Theorem \ref{thm:duhamel}, and as such converges strongly
to its limit. 

In addition to the assumptions on the Green's function in
\eqref{eq:regG} and \eqref{eq:modulusG}, we also assume
that $\rho(x)\in L^1(\Rm^d)$.  Then we have
\begin{theorem}
  \label{thm:conv}
  Let $u_\eps(t,x)$ be the solution constructed in
  Theorem \ref{thm:duhameleps} and $u(t,x)$ the solution constructed in
  Theorem \ref{thm:duhamel}. Then we have that $u_\eps(t,x)$ converges in
  distribution to $u(t,x)$ as $\eps\to 0$. More precisely, let $\tilde
  u_\eps(t,x)$ be the Duhamel solution corresponding to the random
  potential $\tilde q$ in \eqref{eq:tildeq}. Then we have that
  \begin{equation}
    \label{eq:strongconv}
      \|\tilde u_\eps(t)-u(t)\|_{L^2(\Rm^d\times\Omega)} \to 0,\qquad
   \eps\to0,
  \end{equation}
  uniformly in time over compact intervals.
\end{theorem}
\begin{proof}
  Let us drop the upper $\tilde{}$ to simplify notation. We have
  \begin{displaymath}
    \begin{array}{rcl}
    \delta I_\eps(t)&=&\dint_{\Rm^d} \E\{(u(t)-u_\eps(t))^2\} dx 
   =  \dsum_{n,m} \delta I_{\eps,n,m}(t) \\
    \delta I_{\eps,n,m}(t) &=& \dint_{\Rm^d}
     \E\{(u_n(t)-u_{n,\eps}(t))(u_m(t)-u_{m,\eps}(t))\}dx.
    \end{array}
  \end{displaymath}
  Following the proofs of Theorems \ref{thm:duhamel} and
  \ref{thm:duhameleps}, we observe that
  \begin{displaymath}
    \begin{array}{l}
        \delta I_{\eps,n,m}(t) = \\
  \qquad \dint_{\Rm^d} \dprod_{k=0}^{n-1}\dint_0^{t_k}\dint_{\Rm^{dn}}
   \prod_{k=0}^{n-1} G(t_k-t_{k+1},x_k;x_{k+1})
    \dint_{\Rm^d} G(t_n,x_n;\xi) u_0(\xi) d\xi
   \dprod_{k=1}^n dt_k\\
   \qquad
   \dprod_{l=0}^{m-1}\dint_0^{s_l}\dint_{\Rm^{dm}}
   \prod_{l=0}^{m-1} G(s_l-s_{l+1},y_l;y_{l+1})
   \dint_{\Rm^d} G(s_m,y_m;\zeta) u_0(\zeta) d\zeta
   \dprod_{l=1}^m ds_l\delta(x_0-x)\\
   \qquad \delta(y_0-x)
  \E\Big\{\Big(\dprod_{k=1}^n \sigma dW(x_k)-\prod_{k=1}^n q_\eps(x_k)dx_k\Big)
  \Big(\prod_{l=1}^m \sigma dW(y_k)-\prod_{l=1}^m q_\eps(y_l)dy_l\Big)\Big\}dx.
   \end{array}
  \end{displaymath}
  Here, we have again that $t_0=s_0=t$. The integration in $x$ is handled
  as in  the proof of Theorem \ref{thm:duhamel} so that
  \begin{displaymath}
    \begin{array}{l}
        |\delta I_{\eps,n,m}(t)|\lesssim \\
  \qquad \Big| \dint_0^t\phi(t_1)   
     \dprod_{k=1}^{n-1}\dint_0^{t_k}\dint_{\Rm^{d(n-1)}}
   \prod_{k=1}^{n-1} G(t_k-t_{k+1},x_k;x_{k+1})
    \dint_{\Rm^d} G(t_n,x_n;\xi) u_0(\xi) d\xi
   \dprod_{k=1}^n dt_k\\
   \qquad
   \dint_0^t \dprod_{l=1}^{m-1}\dint_0^{s_l}\dint_{\Rm^{d(m-1)}}
   \prod_{l=1}^{m-1} G(s_l-s_{l+1},y_l;y_{l+1})
   \dint_{\Rm^d} G(s_m,y_m;\zeta) u_0(\zeta) d\zeta
   \dprod_{l=1}^m ds_l\\
   \qquad 
  \E\Big\{\Big(\dprod_{k=1}^n \sigma dW(x_k)-\prod_{k=1}^n q_\eps(x_k)dx_k\Big)
  \Big(\prod_{l=1}^m \sigma dW(y_k)-\prod_{l=1}^m q_\eps(y_l)dy_l\Big)\Big\}
   \Big|.
   \end{array}
  \end{displaymath}
  The main difference with respect to previous proofs is that we cannot
  bound the Green's functions by their absolute values just yet.
  The moment of Gaussian variables is handled as follows.
  We recast it as 
  \begin{displaymath}
    \Big(\prod_{k=1}^n \sigma dW(x_k)-\prod_{k=1}^n q_\eps(x_k)dx_k\Big)
         \prod_{l=1}^m q_\eps(y_l)dy_l
  \end{displaymath}
  plus a second contribution that is handled similarly. 
  We denote by $\delta I^1_{\eps,n,m}(t)$ the corresponding contribution
  in $\delta I_{\eps,n,m}(t)$ and by 
  $\delta I^2_{\eps,n,m}(t)=\delta I_{\eps,n,m}(t)-\delta I^1_{\eps,n,m}(t)$.
  The above contribution is recast as
  \begin{equation}\label{eq:contsum}
     \dsum_{q=1}^n\prod_{p=1}^{q-1}\sigma dW(x_k)
    \Big(\sigma dW(x_q)-q_\eps(x_q)dx_q\Big)
     \prod_{p=q+1}^{n+m} q_\eps(x_p)dx_p,
  \end{equation}
  where we have defined $x_{n+l}=y_l$ for $1\leq l\leq m$. We have
  therefore $n$ (or more precisely $n\wedge m$ by decomposing the
  product over $m$ variables when $m<n$) terms of the form
  \begin{displaymath}
    \E\Big\{ \prod_{p=1}^{q-1}\sigma dW(x_k)
    \Big(\sigma dW(x_q)-q_\eps(x_q)dx_q\Big)
     \prod_{p=q+1}^{n+m} q_\eps(x_p)dx_p\Big\} :=
   \E\Big\{ \prod_{k=1}^{2\ban} a_{k,\eps}(dx_k)\Big\},
  \end{displaymath}
  where each measure $a_{k,\eps}(dx_k)$ is Gaussian. Then,
  \eqref{eq:mtGauss} is replaced in this context by
   \begin{equation}
  \label{eq:mtGaussa}
   \begin{array}{rcl}
  \E\Big\{ \dprod_{k=1}^{2\ban} a_{k,\eps}(dx_k)\Big\}&=& \dsum_{\pp\in\PP}
   \dprod_{k\in A_0(\pp)}  \E\big\{a_{k,\eps}(dx_k)
     a_{l(k),\eps}(dx_{l(k)}) \big\}\\
    & :=& \dsum_{\pp\in\PP}
   \dprod_{k\in A_0(\pp)} h_{\eps,k}(x_k-x_{l(k)}) dx_k dx_{l(k)}.
   \end{array}
  \end{equation}
  The functions $h_{\eps,k}(x_k-x_{l(k)})$ come in five different
  forms according as
  \begin{equation}
    \label{eq:formsaeps}
    \begin{array}{rcl}
     \E\{dW(x)dW(y)\} &=& \sigma^2 \delta(x-y)dx dy\\[2mm]
     \E\{dW(x)q_\eps(y)dy\} &=& \sigma \dfrac{1}{\eps^d}
    \rho\Big(\dfrac{x-y}\eps\Big) dx dy \\[2mm]
    \E\{q_\eps(x)dx q_\eps(y) dy \} &=&  \dfrac{1}{\eps^d}
    R\Big(\dfrac{x-y}\eps\Big) dx dy \\[2mm]
    \E\{(dW(x)-q_\eps(x)dx)dW(y)\} &=&
     \Big(\sigma^2 \delta(x-y) - \sigma \dfrac{1}{\eps^d}
    \rho\Big(\dfrac{x-y}\eps\Big) \Big)  dx dy \\[2mm]
    \E\{(dW(x)-q_\eps(x)dx)q_\eps(y)dy\} &=&
    \Big(\sigma \dfrac{1}{\eps^d}
    \rho\Big(\dfrac{x-y}\eps\Big) - \dfrac{1}{\eps^d}
    R\Big(\dfrac{x-y}\eps\Big)\Big)  dx dy.
    \end{array}
  \end{equation}
  At this point, we have obtained that 
 \begin{displaymath}
    \begin{array}{l}
        |\delta I^1_{\eps,n,m}(t)|\lesssim 
   \dsum_{\pp\in\PP} \\
  \qquad \Big| \dint_0^t \phi(t_1)
   \dprod_{k=1}^{n-1}\dint_0^{t_k}\dint_{\Rm^{d(n-1)}}
   \prod_{k=1}^{n-1} G(t_k-t_{k+1},x_k;x_{k+1})
    \dint_{\Rm^d} G(t_n,x_n;\xi) u_0(\xi) d\xi
   \dprod_{k=1}^n dt_k\\
   \qquad
   \dint_0^t \dprod_{l=1}^{m-1}\dint_0^{s_l}\dint_{\Rm^{d(m-1)}}
   \prod_{l=1}^{m-1} G(s_l-s_{l+1},y_l;y_{l+1})
   \dint_{\Rm^d} G(s_m,y_m;\zeta) u_0(\zeta) d\zeta
   \dprod_{l=1}^m ds_l\\
   \qquad 
   \dprod_{k\in A_0(\pp)} h_{\eps,k}(x_k-x_{l(k)}) dx_k dx_{l(k)}
   \Big|.
   \end{array}
  \end{displaymath} 
  Using the notation as in the proof of Theorem \ref{thm:duhamel}, we
  obtain that
   \begin{displaymath}
    \begin{array}{l} |\delta I^1_{\eps,n,m}(t)|\lesssim \dsum_{\pp\in\PP} \Big|
    \dint_{\Rm^{2d}} \dint_0^t\phi(t_1) \dprod_{k=1}^{n-1}\dint_0^{t_k}
    \dint_0^t \dprod_{l=1}^{m-1}\dint_0^{t_{n+l}}
    u_0(\xi)u_0(\zeta)\dint_{\Rm^{\ban d}}\dprod_{k\in A_0(\pp)}  \\
       \Big(
     G(t_k-\tau_k,x_k;y_k)G(t_{l(k)}-\tau_{l(k)},x_{l(k)};y_{l(k)})
     h_{\eps,k}(x_k-x_{l(k)}) dx_k dx_{l(k)}\Big)
    d\xi d\zeta \dprod_{k=1}^{n+m} dt_k \Big|\\
    \lesssim  \dsum_{\pp\in\PP}  
     \dint_{\Rm^{2d}} \dint_0^t\phi(t_1) \dprod_{k=1}^{n-1}\dint_0^{t_k}
    \dint_0^t \dprod_{l=1}^{m-1}\dint_0^{t_{n+l}}
     |u_0(\xi)||u_0(\zeta)| \Big| \dint_{\Rm^{\ban d}}\dprod_{k\in A_0(\pp)}
      \\
       \Big(
     G(t_k-\tau_k,x_k;y_k)G(t_{l(k)}-\tau_{l(k)},x_{l(k)};y_{l(k)})
     h_{\eps,k}(x_k-x_{l(k)}) dx_k dx_{l(k)}\Big) \Big| 
     d\xi d\zeta \dprod_{k=1}^{n+m} dt_k.
   \end{array}
  \end{displaymath}
  It remains to handle the multiple integral between absolute values.
  For $k\in A_0(\pp)$ for which $h_{\eps,k}$ is of the form given in
  the last two lines of \eqref{eq:formsaeps}, we observe that the
  corresponding term between parentheses in the above expression is of
  the form
  \begin{equation}
    \label{eq:formh}
    \begin{array}{rcl}
           && \dint_{\Rm^{2d}} G(s,x;\zeta)G(\tau,y;\xi) h_\eps(x-y)dxdy\\[4mm]
   &=& \dint_{\Rm^d} G(s,x,\zeta) \Big(
    \dint_{\Rm^d} \dfrac{1}{\eps^{d}} g\Big(\dfrac{x-y}\eps\Big)
     \big(G(\tau,x;\xi)-G(\tau,y;\xi)\big) dy\Big)dx\\[4mm]
   &=& \dint_{\Rm^d} G(s,x,\zeta) \Big(  \dint_{\Rm^d} 
        g(y) \big(G(\tau,x;\xi)-G(\tau,x+\eps y;\xi)\big) dy\Big)dx,
    \end{array}
  \end{equation}
  where the function $g(x)$ is given by 
  \begin{equation}
    \label{eq:fctg}
    g(x) = \pm \sigma \rho(x), \qquad \mbox{ or } \qquad
    g(x) = \pm (R(x)-\sigma \rho(x)).
  \end{equation}
  This is because $\rho$ averages to $\sigma$ while $R$ averages to
  $\sigma^2$. Let $k_0$ be the index for which $h_{\eps,k_0}$ is in
  the form of a difference as above. This yields, with $g=g[k_0]$ as
  above,
 \begin{displaymath}
    \begin{array}{l} |\delta I^1_{\eps,n,m}(t)|\lesssim \dsum_{\pp\in\PP}  
     \dint_{\Rm^{2d}} \dint_0^t\phi(t_1) \dprod_{k=1}^{n-1}\dint_0^{t_k}
    \dint_0^t \dprod_{l=1}^{m-1}\dint_0^{t_{n+l}}
     |u_0(\xi)||u_0(\zeta)| \dint_{\Rm^{\ban d}}\dprod_{k_0\not=k\in A_0(\pp)}
      \\
       \Big|
     G(t_k-\tau_k,x_k;y_k)G(t_{l(k)}-\tau_{l(k)},x_{l(k)};y_{l(k)})
     h_{\eps,k}(x_k-x_{l(k)})\Big| dx_k dx_{l(k)}\\
     |G(t_{\mk(k_0)}-\tau_{\mk(k_0)},x_{\mk(k_0)};y_{\mk(k_0)})|
     |g(x_{\mk'(k_0)})|d\xi d\zeta \dprod_{k-1}^{n+m} dt_k\\
  \Big|\big(G(\cdot,\cdot,\cdot)-G(\cdot,\cdot+\eps x_{\mk'(k_0)},\cdot)\big)
      (t_{\mk'(k_0)}-\tau_{\mk'(k_0)},x_{\mk(k_0)};y_{\mk'(k_0)})\Big|
     dx_{k_0}dx_{l(k_0)}.
   \end{array}
  \end{displaymath}
  The above term is now handled as in the proof of Theorem
  \ref{thm:duhamel}. For $k\not=k_0$, the bounds are obtained as
  before because $\rho$ and $R$ are integrable functions by
  hypothesis. The Green's function
  $|G(t_{\mk(k_0)}-\tau_{\mk(k_0)},x_{\mk(k_0)};y_{\mk(k_0)})|$ is
  bounded by a constant times
  $|t_{\mk(k_0)}-\tau_{\mk(k_0)}|^{-\alpha}$. The integration
  $dx_{k_0}dx_{l(k_0)}=dx_{\mk(k_0)}dx_{\mk'(k_0)}$ then yields a
  contribution bounded by $|t_{\mk'(k_0)}-\tau_{\mk'(k_0)}|^{-\gamma}$
  times
  \begin{equation}
    \label{eq:Meps}
    M_\eps = \sup\limits_{\tau\in(0,T),\xi\in\Rm^d}\tau^\gamma
    \dint_{\Rm^{2d}} |g(y)| \big|G(\tau,x;\xi)-G(\tau,x+\eps y;\xi)\big|
    dx dy.
  \end{equation}
  The presence of the factor $\gamma$ is necessary in order for
  $M_\eps$ to converge to $0$ as $\eps\to0$. As a consequence, as in
  the derivation of \eqref{eq:Imn1}, we obtain that
  \begin{displaymath}
    |\delta I_{\eps,n,m}(t)| 
    \lesssim \dsum_{\pp\in\PP} \dint_0^t\dprod_{k=1}^{n-1}\dint_0^{t_k}
   \dint_0^t\dprod_{l=1}^{m-1}\dint_0^{t_{n+l}}
    \dfrac{2n M_\eps\phi(t_1)}{|t_{\mk'(k_0)}-\tau_{\mk'(k_0)}|^{\gamma}}
    \dprod_{k\in A_0}
    \dfrac{C}{(t_{\mk(k)}-\tau_{\mk(k)})^{\frac d\m}}
   \!\! \dprod_{k=1}^{n+m}\!\! dt_k.
  \end{displaymath}
  The factor $2n$ comes from twice the summed contributions in
  \eqref{eq:contsum}. The presence of the factor $\gamma$ increases
  the time integrals as follows. Assume that $k_0\leq n$ for
  concreteness; the case $k_0\geq n+1$ is handled similarly. Then
  $\beta_0$ in the proof of Theorem \ref{thm:duhamel} should be
  replaced by $\beta_0+\gamma$.  This does not significantly modify
  the analysis of the $\Gamma$ functions and the contribution of each
  graph is still bounded by a term of the form \eqref{eq:bdnm}. The
  behavior in time, however, is modified by the presence of the
  contribution $\gamma$ and we find that
  \begin{displaymath}
    |\delta I_{\eps,n,m}(t)| \leq C n  M_\eps 
       t_0^{(n+m)(1-\frac\alpha2)-\alpha-\gamma}  C^n C^m 
   \dfrac{1}{n^{\frac n2(1-\alpha)}m^{\frac m2(1-\alpha)}}.
  \end{displaymath}
  The above bound is of interest for $n+m\geq2$ since the case $n=m=0$
  corresponds to the ballistic component $u_0(t,x)$, which is the same
  for $u_\eps(t,x)$ and $u(t,x)$ so that $\delta I_{\eps,0,0}=0$.  By
  choosing $\gamma=2(1-\alpha)>0$, we observe that
  $2(1-\frac\alpha2)-\alpha-\gamma\geq0$ for $n+m\geq2$ so that $|\delta
  I_{\eps,n,m}(t)|$ is bounded uniformly in time. The new factor $n$
  may be absorbed into $C^n$ so that after summation over $n$ and $m$,
  we get
  \begin{equation}
    \label{eq:bounduueps}
    \|u_\eps(t)-u(t)\|^2_{L^2(\Rm^d\times\Omega)}
     \leq C M_\eps.
  \end{equation}
  By assumption \eqref{eq:modulusG}, the integrand in \eqref{eq:Meps}
  converges point-wise to $0$ and an application of the dominated
  Lebesgue convergence theorem shows that $M_\eps\to0$. This concludes
  the proof of the convergence result.
\end{proof}

\paragraph{A continuity lemma.} We conclude this paper by showing that 
the constraints \eqref{eq:regG} and \eqref{eq:modulusG} imposed on the
Green's functions of the unperturbed problem throughout the paper are
satisfied for a natural class of parabolic operators.
\begin{lemma}
  \label{lem:bounds}
  Let $G(t,x)$ be defined as the Fourier transform of
  $e^{-t|\xi|^\m}$, i.e., the Green's function of the operator
  $\partial_t+(-\Delta)^{\frac \m2}$ for $\m>d$. Then the conditions
  in \eqref{eq:regG} and \eqref{eq:modulusG} are satisfied.  Moreover,
  when $\m$ is an even number, then $M_\eps$ in \eqref{eq:Meps}
  satisfies the bound
  \begin{displaymath}
    M_\eps \lesssim \eps^\beta, \qquad \beta= 2(\m-d)\wedge 1.
  \end{displaymath}
\end{lemma}
\begin{proof}
  By scaling, we find that $G(t,x)=t^{-\frac d\m}G(1,t^{-\frac
    1\m}x)$. Since $|\xi|^pe^{-|\xi|^\m}$ is integrable for all $p$,
  we obtain that $G(1,x)$ belongs to $C^\infty(\Rm^d)$. Since $G(1,x)$
  is bounded, then so is $t^{\frac d\m}G(t,x)$ uniformly in $t$ and
  $x$. 
  
  By the above scaling, $G(t,x)$ belongs to $L^1(\Rm^d)$ uniformly in
  time if and only if $G(1,x)$ does.  When $\m$ in an even integer,
  then $e^{-|\xi|^\m}$ belongs to $\mathcal S(\Rm^d)$, the space of
  Schwartz functions, so that $G(1,x)\in \mathcal S(\Rm^d)$ as well.
  It is therefore integrable and has an integrable gradient.  When
  $\m$ is not an even integer, we have
  \begin{displaymath}
    e^{-|\xi|^\m}-1 = \dsum_{k=1}^\infty \dfrac{(-1)^k}{k!}
     |\xi|^{k\m}.
  \end{displaymath}
  The Fourier transform of the homogeneous function $|\xi|^{k\m}$ 
  is given by \cite{Taylor-PDE-1}
  \begin{displaymath}
    c(k) |x|^{-k\m-d}, \qquad c(k) = C_d 2^{k\m} \dfrac{\Gamma(\frac12(k\m+d))}
   {\Gamma(-\frac12 k\m\,)},
  \end{displaymath}
  where $C_d$ is a normalization constant independent of $k$. The
  Fourier transform of $e^{-|\xi|^\m}$ may then be written as a
  constant times $|x|^{-(d+\m)}$ plus a smoother contribution that
  converges faster to $0$ (for instance because it belongs to some
  $H^s(\Rm^d)$ with $s>\frac d2+k$ sufficiently large so that $k$
  derivatives of this contribution are integrable). It is therefore
  integrable for $\m>0$. The $L^2$ bound follows from the $L^1$ and
  $L^\infty$ bounds.
  
  We obtain by scaling and from definition of $G(t,x)$ that
  \begin{displaymath}
    \begin{array}{ll}
    &t^\gamma\dint_{\Rm^d} |G(t,x)-G(t,x+\eps y)| dx =
   t^\gamma\dint_{\Rm^d} |G(1,x)-G(1,x+t^{-\frac 1\m}\eps y)| dx .
    \end{array}
  \end{displaymath}
  The above derivation shows that the gradient of $G$ is also
  integrable for $\m>1$ so we may bound the above quantity by
  $t^\gamma(1\wedge t^{-\frac 1\m}\eps |y|)$.  Now,
  \begin{displaymath}
    \sup_{t<T} (t^\gamma\wedge t^{\gamma-\frac 1\m} \eps|y|)
    \lesssim (\eps|y|)^{\gamma \m}\vee \eps|y|=(\eps|y|)^{2(\m-d)}
    \vee \eps|y|,
  \end{displaymath}
  according as $\gamma\m<1$ or $\gamma\m\geq1$. We thus obtain
  \eqref{eq:modulusG} by sending $\eps y\to0$.  When $g(y)$ is
  sufficiently regular, then we obtain the more precise bound
  \begin{displaymath}
     M_\eps \lesssim \eps^{2(\m-d)}
     \dint_{\Rm^d} |g(y)| |y|^{2(\m-d)} dy
    \vee \eps \dint_{\Rm^d} |g(y)| |y| dy,
  \end{displaymath}
  provided that the latter integrals are well-defined.
\end{proof}

%%%%%%%%%%%%%%%%%%%%%%%%%%%%%%%%%%%%%%%%%%%%%%%%%%%%%%%%%%%%%%%%%%
%%%%%%%%%%%%%%%%%%%%%%%%%%%%%%%%%%%%%%%%%%%%%%%%%%%%%%%%%%%%%%%%%%
\section*{Acknowledgment}
%%%%%%%%%%%%%%%%%%%%%%%%%%%%%%%%%%%%%%%%%%%%%%%%%%%%%%%%%%%%%%%%%%
%%%%%%%%%%%%%%%%%%%%%%%%%%%%%%%%%%%%%%%%%%%%%%%%%%%%%%%%%%%%%%%%%%

This work was supported in part by NSF Grants DMS-0239097 and
DMS-0804696.

%%%%%%%%%%%%%%%%%%%%%%%%%%%%%%%%%%%%%%%%%%%%
%%%%%%%%%%%%%%%%%%%%%%%%%%%%%%%%%%%%%%%%%%%%
\bibliography{../../bibliography} \bibliographystyle{siam}
%%%%%%%%%%%%%%%%%%%%%%%%%%%%%%%%%%%%%%%%%%%%
%%%%%%%%%%%%%%%%%%%%%%%%%%%%%%%%%%%%%%%%%%%%

%%%%%%%%%%%%%%%%%%%%%%%%%%%%%%%%%%%%%%%%%%%%
\end{document}